\documentclass{fmj2010}
\usepackage{graphicx}
\begin{document}

\title{Cross-Analyzing Radio and $\gamma$-Ray Time Series Data: \\Fermi Marries Jansky }

\author{Jeffrey D. Scargle\thanks{On behalf of the Fermi/LAT Collaboration.
   This work is funded by the NASA
   Applied Information Sciences Research Program
   and a Fermi Guest Investigator
   grant with Jay Norris, James Chiang and 
Roger Blandford as
co-Investigators.}}

\institute{Space Science Division,
   NASA-Ames, Moffett Field, CA  94035, USA;
  Fermi Gamma Ray Space Telescope}
  
%\authorrunning{J. Scargle}
%\titlerunning{Cross-Analyzing Radio and Gamma-Ray Time Series Data}

 \abstract{A key goal of radio and $\gamma-$ray
observations of active galactic nuclei 
is to characterize their time variability 
in order to elucidate physical processes
responsible for the radiation.   
I describe algorithms for relevant 
time series analysis tools -- 
correlation functions, 
Fourier and wavelet amplitude and phase spectra, 
structure functions, and time-frequency distributions,
all for arbitrary data modes and sampling schemes.
For example radio measurements 
can be cross-analyzed with data streams
consisting of time-tagged gamma-ray photons. 
Underlying these methods is the Bayesian block
scheme, useful in its own right to characterize local
structure in the light curves, and also prepare 
raw data for input to the other analysis
algorithms.
One goal of this presentation 
is to stimulate discussion of these
methods during the workshop.}

\maketitle

%\authorrunning{J. Scargle}
%\titlerunning{Cross-Analyzing Radio and Gamma-RayTime Series Data}
%\maketitle
%
%______________________________________________________________
\section{Introduction}
%______________________________________________________________

Active galactic nuclei (AGN) are highly variable
at all wavelengths.  
A major fraction of their total luminosity 
fluctuates over time scales ranging from the
shortest for which statistically significant
signals can be obtained, to the longest 
time intervals over which data are available.
Characterizing this variability 
has yielded growing insight into the physical 
processes powering the large AGN
luminosities -- a trend that will accelerate
as observations, data analysis, and 
theory proliferate.

This paper outlines time
series methods %, both standard and novel,
for analysis of the disparate data modes
of radio, $\gamma-$ray, and other astronomical observations.
The next section introduces a data structure
that generalizes data modes 
traditionally used  for time-sequential observations.
This abstraction yields 
methods for estimating,
from arbitrary time series data,
including heterogeneous 
mixtures of data modes,
all of the
standard analysis functions:
\begin{description}
\item[$\bullet$] light curves
\item[$\bullet$] autocorrelations
\item[$\bullet$] Fourier power and phase spectra
\item[$\bullet$] wavelet representations 
\item[$\bullet$] structure functions
\item[$\bullet$] time-frequency distributions
\end{description}
\noindent
As indicated in Table 1,
for essentially arbitrary data modes
these methods yield amplitude and phase information
for single or multiple time series (auto- and cross- analysis,
respectively) -- if desired, conditional on auxiliary variables.

%========================================
\begin{table}[htdp]
\caption[]{Time Series Analysis for All Data Modes}
\label{tsaf_table}
 \vskip -.75cm
 $$ 
         \begin{tabular}{l|ccccc}
           \hline
           % \noalign{\smallskip}
            \  & Auto  &Cross &Amp &Phase &Condit. \\
            % \noalign{\smallskip}
            \hline
            \noalign{\smallskip}
            Correlation & yes  & yes  & - & - & yes \\
            Fourier &  yes   & yes    & yes & yes & yes \\
            Wavelet  & yes & yes & yes & location  & ? \\
            Struct Fcn &  yes   & yes   & - & - &  yes     \\
            Time-Freq  &  yes   & yes & yes & yes & yes \\
            \noalign{\smallskip}
            \hline
         \end{tabular}
$$ 

\end{table}

%========================================

There are significant difficulties 
in the astrophysical
interpretation of these quantities.
The methods described here
are of use in some of these,
such as
separation of 
observational errors from stochastic 
source variability (both of which,
unfortunately, are often called \emph{noise}).
But I do not discuss other
more difficult problems,
which are probably beyond the scope of
time series analysis methods,
such as assessing the importance 
of \emph{cosmic variance},
identifying causal or otherwise 
physically connected relationships 
in multi-wavelength time series data, \emph{etc}.

Subsequent sections discuss each of the
above-listed functions 
and give sample applications.

%------------------------------------------------------
\section{Abstract Data Cells}
%------------------------------------------------------

The time series algorithms 
to be described below 
can be applied to
almost any type of time-sequential 
astronomical data.
This generality is facilitated by 
identifying those features of the data modes 
that are necessary for analysis algorithms.

Each individual act of measurement
may yield a large set of data values
relevant to estimation of the signal
amplitude, and its uncertainty, as
a function of time.
Of these, two pieces of information,
related to the independent variable
(time\footnote{In practice 
we always use a
discrete time representation,
such as a micro-second scale
computer clock tick, or the finite time
interval of signal averaging. } 
of the measurement)
and the dependent variable
(amplitude of the signal at that time), 
are necessary for 
any time series algorithm.
In radio astronomy the typical example
is the measurement of the flux
of a source averaged over a 
short interval of time. 
In $\gamma$-ray
astronomy the typical example
is the detection of individual photons.
The arrival time of the photon is
obviously the timing quantity,
but what about the signal?
One scheme is to represent
an individual photon with 
a delta-function in time. 
But more information can 
be extracted by incorporating
the time intervals\footnote{A method
for analyzing event data based solely
on inter-event time intervals has
been developed in (\cite{prahl}).}
between photons.
Specifically, for each photon
consider the interval starting half way
back to the previous photon and ending 
half way forward to the subsequent photon.
This interval, namely 
\begin{equation}\label{interval}
[ {t_{n} - t_{n-1} \over 2}, {t_{n+1} - t_{n} \over 2} ]\ , 
\end{equation}
\noindent
is the set of times closer to $t_{n}$ than
to any other time,\footnote{These intervals form the
{\emph Voronoi tessellation} of the
total observation interval. See (\cite{spatial_tessellations})
for a full discussion of this construct,
highly useful in spatial domains of 2, 3, or
higher dimension; see also (\cite{scargle_2,scargle_4}).}
and has length equal to the average of the
two intervals connected by photon $n$, namely 
\begin{equation}
 \Delta t_{n} = { t_{n+1} - t_{n-1} \over 2} \ .
\end{equation}
Then the reciprocal 
\begin{equation}\label{reciprocal}
x_{n} \equiv { 1 \over \Delta t_{n} }
\end{equation}
is taken as an estimate
of the signal amplitude corresponding 
to observation $n$.
When the photon rate 
is large, the corresponding intervals are small.
Figure 1
\begin{figure}
\centering
\vspace{40pt}
\includegraphics[width=8cm]{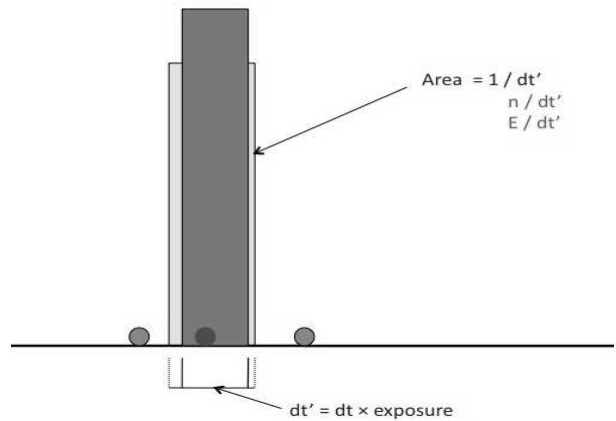}
\caption{Voronoi cell of a photon.
Three successive photon detection times are 
circles on the time axis.  The vertical
dotted lines underneath delineate the time cell, 
and the light rectangle is the local estimate of the
signal amplitude.  If the exposure at this
time is less than unity, the length of the
data cell shrinks in proportion ($dt \rightarrow dt'$),
yielding a larger estimate of the true event rate 
(darker rectangle).
The height of the rectangle is ${n / dt'}$,
where $n$ is the number of photons at exactly 
the same time (almost always 1), 
or by the photon energy for a flux estimate.}
\label{interval}
\end{figure}
demonstrates the data cell concept,
including the simple modifications to
account for variable exposure 
and for weighting
by photon energy.

Consider gaps in the data.
By this we mean that there are
portions of the total observation interval
during which the detection system
is completely off (exposure zero).
This situation is readily handled
by defining the start of the data cell 
for the first photon detected after the
gap at the end time of the gap.
Correspondingly the data cell for
the last photon before the gap 
is set at the start time of the gap.
The statistical nature of independent 
events assures that this procedure
rigorously estimates the true photon
rate at the edge of the gap.
Of course, no information is
available about the signal during
such a gap, and 
the various algorithms deal with
gaps accordingly.

Now we consider data modes generally.
Three common examples are: 
(a) measurements of a quasi-continuous physical variable
(eg. radio astronomy flux measurements)
(b) the time of occurrence of discrete events
(e.g. photons)
and
(c) counts of events in bins.
The signal of interest is
the time dependence 
of the measured quantity in case (a), 
or of the event rate in case (b).
Case (c) is actually very similar to (b), but
is often described as density
estimation or determination of 
the event distribution function.
In all cases it is useful to 
introduce the concept of 
\emph{cells}
to represent the measurements. 
Letting ${\bf x_{n}}$ be the estimate of the 
signal amplitude for a cell at time $t_{n}$, 
a data set of $N$ sequential observations 
is denoted
\begin{equation}
C_{n} \equiv \{ {\bf x}_{n}, t_{n} \} , 
 \ \ \  n = 1, 2, \dots, N .
\label{data_intro}
\end{equation}
\noindent The specific meaning of the quantities ${\bf x}_{n}$ 
depends on the type of data.
For example, in the three cases mentioned above
the array ${\bf x}$  contains 
(a) the sum or average
over the measurement interval
of an extensive or intensive quantity,
respectively,
plus one or more quantifiers of 
measurement uncertainty, 
(b) coordinates of events, 
such as photon arrival times,
and
(c) sizes and locations of the bins,
and the count of events in them.

A major reason for constructing this
abstract data representation is that
it unifies all data modes into a common
format that makes construction of
universal algorithms easy.
As we will see in the next
sections, even  mixtures of
data types -- either in the sense
of cross-analyzing two very different
data types, or mixing data within
a single time series -- can be handled.

%__________________________________________________________
\section{Light Curve Analysis: Bayesian Blocks}

The simplest and most direct way to 
study variability is to construct a representation 
of the intensity of the source as a function of time.
More can be done than 
just plotting the intensity measurements 
as a function of the time of the measurement.
Smoothing, interpolation, gap filling,
etc. are all techniques meant to enhance
one's understanding of the variability.
Here we discuss a different procedure, 
namely construction of a simple,
generic, non-parametric
model of the data that as 
much as possible shows 
the actual variability of the source, 
and minimizes the effect of observation errors.
The model adopted is the simplest possible
non-parametric representation of 
time series data, namely a piece-wise
constant model.
Details of this approach are
given in (\cite{scargle_new});
the improved algorithm 
given there replaces the
approximate one described in
(\cite{scargle_v}).

The Bayesian Blocks algorithm 
finds the best partition of the observation
interval into blocks, such that the 
source intensity is modeled as 
varying from block to block, 
but constant within each block.
This is just a step-function representation
of the data.  
The meaning of the ``best'' model is
the one that maximizes a measure of
goodness-of-fit function described in detail
in (\cite{scargle_new}).
Another change since the earlier
reference is the use of a very simple
maximum likelihood fitness function,
preferable to the Bayesian posterior
previously used because it is 
invariant to a scale change in the
time variable, thus eliminating 
a parameter from the analysis.

Figure 2 shows the Bayesian 
\begin{figure}
\centering
\vspace{40pt}
\label{bb_fig}
%\special{psfile=fig_1.ps hscale=100 vscale=70100 hoffset=20 voffset=0}
\includegraphics[width=8cm]{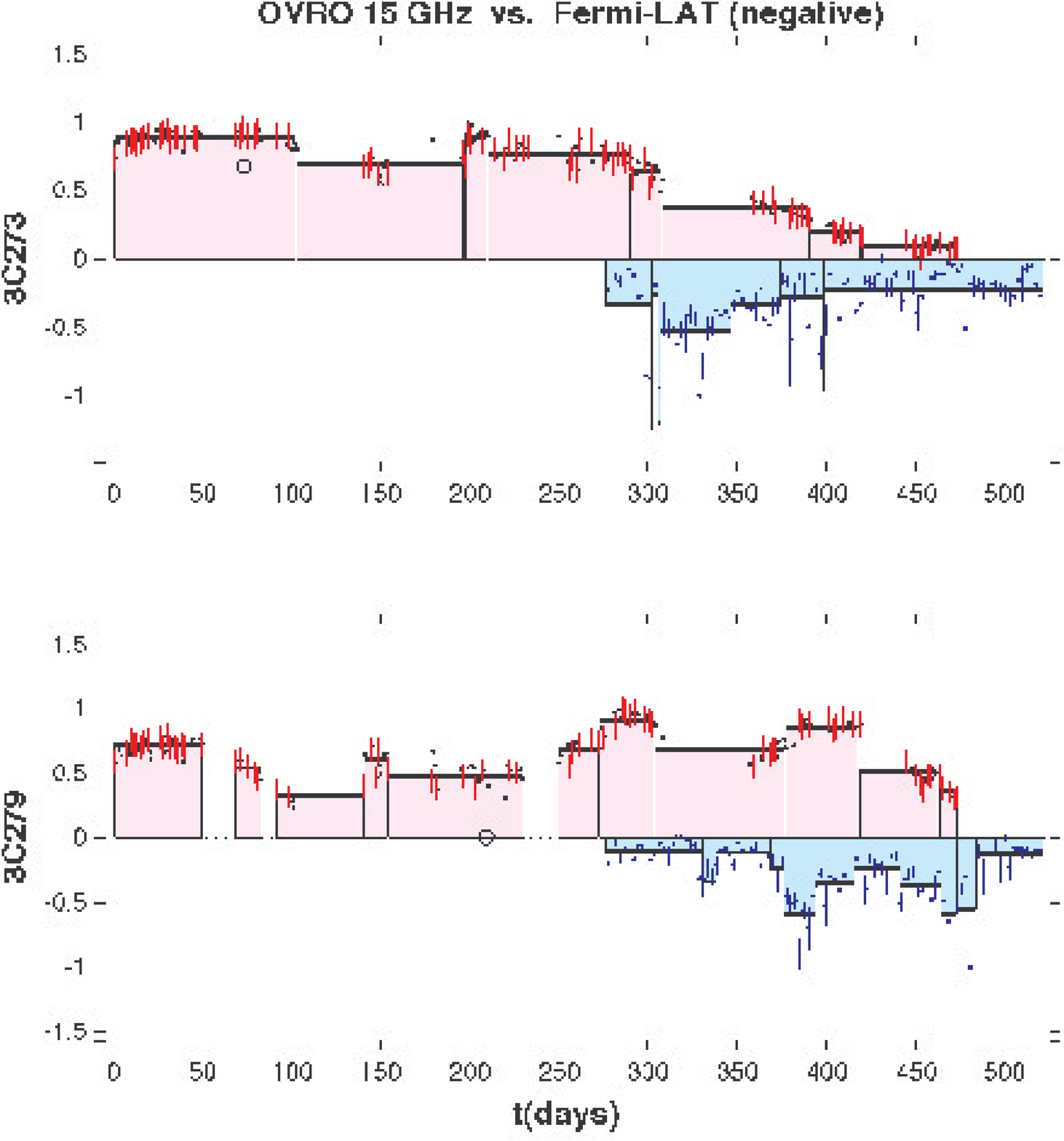}
\caption{Bayesian Block representations of
the lightcurves of 3C273 and 3C279, two 
AGN in the OVRO/Fermi project.  
The co-aligned times are in days,
relative to an arbitrary zero point; amplitudes are
on a common relative scale.  Binned LAT data
is shown for comparison, but the BB representation
is based on the photon data only.}
\end{figure}
blocks analysis of two AGN 
in the OVRO/Fermi joint program.
The data shown are from somewhat
earlier in the program, where the 
overlap between the to instruments
was not huge.  Also these were
just the first and third objects in
the long list of observed sources,
and were not particularly selected 
for being highly variable cases.
%__________________________________________________________
\section{Correlation Functions}
\label{auto_correlation}

\begin{figure}
\centering
\vspace{30pt}
%\special{psfile=fig_1.ps hscale=100 vscale=70100 hoffset=20 voffset=0}
\includegraphics[width=8cm]{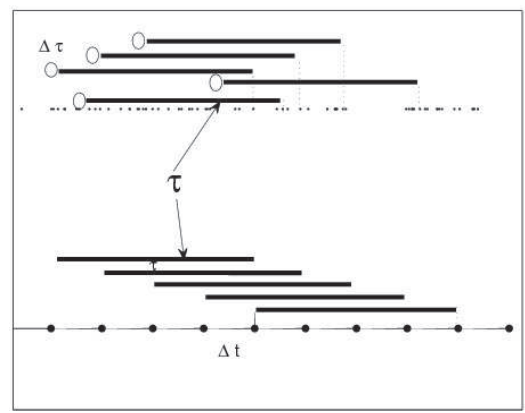}
\caption{Summation schemes for autocorrelation functions.  
The points represent data cells, derived from 
measured values (as in radio astronomy)
or time-tagged events (as in Fermi photon data).  
{\bf Top: }
Summation over data with arbitrary spacing in the Edelson and 
Krolik algorithm. From each point
average over all points within a bin $d\tau$ distant by $\tau$;
$\tau$ is binned, but $t$ is not.
{\bf Bottom: }
Standard lag summation over evenly spaced data. From 
each point (except near the ends) there is another point distant
by exactly $\tau =$ an integer multiple of $\Delta t$.
}
\label{eandk}
\end{figure}

A rather underutilized technique for
studying correlated variability of two
observables (such as time series for 
different wavelengths) is to construct 
a scatter plot of one against the other.
If done carefully, this approach allows
study of joint probability distributions
for the two variables; these contain
more statistical information than correlation
functions or any of the other functions
discussed here.  The challenges of 
this approach include the difficulty 
of depicting the all-important time-sequence
connecting the points in the scatter plot,
and the need to consider plotting 
lagged versions of the variables, for 
a number of values of the lag.
The understanding that comes from
careful study of scatter plots most 
often makes it worthwhile to conquer
these difficulties.

But probably the most used tool for studying 
statistical variability properties of a single
time series is the auto-correlation function (ACF) 
or, for studying relations between
the variability in two or more sets of
simultaneous time series, 
the cross-correlation function (CCF).
The meaning of the latter can be understood
by modeling one time series 
as a lagged version of the other, 
and evaluating the 
posterior distribution of the lag $\tau$, yielding
\begin{equation}
P( \tau ) \sim e^{  R_{X,Y} (\tau )  \over K} \ ,
\end{equation}
where K is a constant and $R_{X,Y} (\tau )$
is the cross-correlation function defined below
(\cite{scargle_3}).

Concentrating on the CCF,
of which the ACF is really a special case,
and following the notation
and definitions of (\cite{papoulis_1,papoulis_2}), 
we have this definition of the
\emph{cross-correlation function} 
of two real processes 
${\bf x}(t)$ and  ${\bf y}(t)$
\begin{equation}
R_{xy}(t_{1}, t_{2} ) = < \{ \  {\bf x}(t_{1} )  {\bf y}(t_{2} ) \   \} >
\end{equation}
\noindent
Assuming the processes are
stationary, the time dependence
is on only the difference 
$\tau \equiv t_{2} - t_{2}$
and we have
\begin{equation}
R_{xy}( \tau ) = < \{ \  {\bf x}(t)  {\bf y}(t + \tau) \   \} >
\end{equation}
\noindent
The symbol $< >$ 
means the \emph{expected value}, 
informally to be thought of as
an average over realizations of the 
underlying random process $X$.
In data analysis 
this theoretical quantity is typically
not known, and must be
therefore be estimated 
from the data at hand, \emph{e.g.}
\begin{equation}
E[ X(t) Y(t+\tau ) ] \equiv  
{1 \over N(\tau)} \sum_{n} x_{n} y_{n+\tau} \
\end{equation}
\noindent
where  $x_{n}$  and $y_{n}$ 
are the samples
of the variable $X, Y$\footnote{{\bf Caution:}
It is 
common to center the processes
about their means, to yield the
\emph{cross-covariance} 
and \emph{auto-covaraince}
functions.
Such mean-removal
can have unfortunate consequences,
such as distortion of the low-frequency
power spectrum.
In addition, the nomenclature is not completely
standard.  Various terms are used
for the cases where the means
of the processes have been subtracted off,
and/or the resulting function normalized
to unity at $\tau = 0$.},
and $N(\tau)$ is the 
number of terms 
for which the sum can be taken.

Figure 3 is a cartoon of the lag relationships
for correlation functions of evenly spaced data (bottom),
as well as a solution to the 
difficulty posed by unevenly
spaced 
time samples in general, 
and event data in particular.  
For a given
sample or event at $t_{n}$ there will in
general not be a corresponding one 
at  $t_{n} + \tau$, no matter what 
restriction is placed on $\tau$.

For this problem
an ingenious if straightforward
algorithm (\cite{edelson_krolik}) is in wide use.
The basic idea is to pre-define 
a set of bins in the variable $\tau$
in order to 
construct a histogram of the 
corresponding time separations
$\tau = t_{m} - t_{n}$, 
weighted
by the corresponding $x_{n} y_{m}$ 
product.
To be more specific, and 
modifying slightly Edelson and Krolik's
formulas for our case (including
not subtracting 
the process means), define 
for all measured pairs $(x_{n}, y_{m})$
the quantity
\begin{equation}\label{e_and_k}
UDFC_{nm} = { x_{n} y_{m} \over 
\sqrt{ ( \sigma_{x} ^{2}  - e_{x}^{2} ) 
( \sigma_{y} ^{2}  - e_{y}^{2} )  } } \ ,
\end{equation}
\noindent
(for Unbinned Discrete Correlation Function) 
where 
$\sigma_{x}$ is the
standard deviation of the
$X$-observations, 
$e_{x}$ is the $X$-measurement
error, and similarly for $Y$.
The estimate of the correlation
function is then
\begin{equation}\label{e_and_k_sum}
R_{xy}( \tau ) = {1 \over N_{\tau} }
\sum UDCF_{nm}
\end{equation}
\noindent
where the sum is over the
pairs, $N_{\tau}$ in number, 
for which $t_{m} - t_{n}$
lies in the corresponding $\tau$-bin.

There has been some confusion
over the rationale for the denominator
in eq. (\ref{e_and_k})
(`` ... to preserve the proper normalization'')
and how to estimate it.
The quantity $( \sigma_{x} ^{2}  - e_{x}^{2} )$
is in principle the difference between the
total observed variance and that ascribed
to observational errors.
How they are estimated from
source and calibration data, and
other instrumental considerations,
no doubt varies from case to case.
Edelson and Krolik discuss 
potential corruption by correlated
observational errors.  
I recommend following their advice
to exclude the terms $n = m$ 
from eq. (\ref{e_and_k_sum})
only for autocorrelations, and
then only if it is really necessary.
These terms yield a 
spike in the autocorrelation
function at $\tau = 0$,
which can be 
a convenient visual assessment of the importance
of the observational variance; it can be easily 
removed if needed.
For CCFs it makes no sense to 
remove these terms, absent observational
errors correlated between the two observables.
\begin{figure}
\centering
\vspace{30pt}
\label{acf_examples}
%\special{psfile=fig_1.ps hscale=100 vscale=70100 hoffset=20 voffset=0}
\includegraphics[width=8cm]{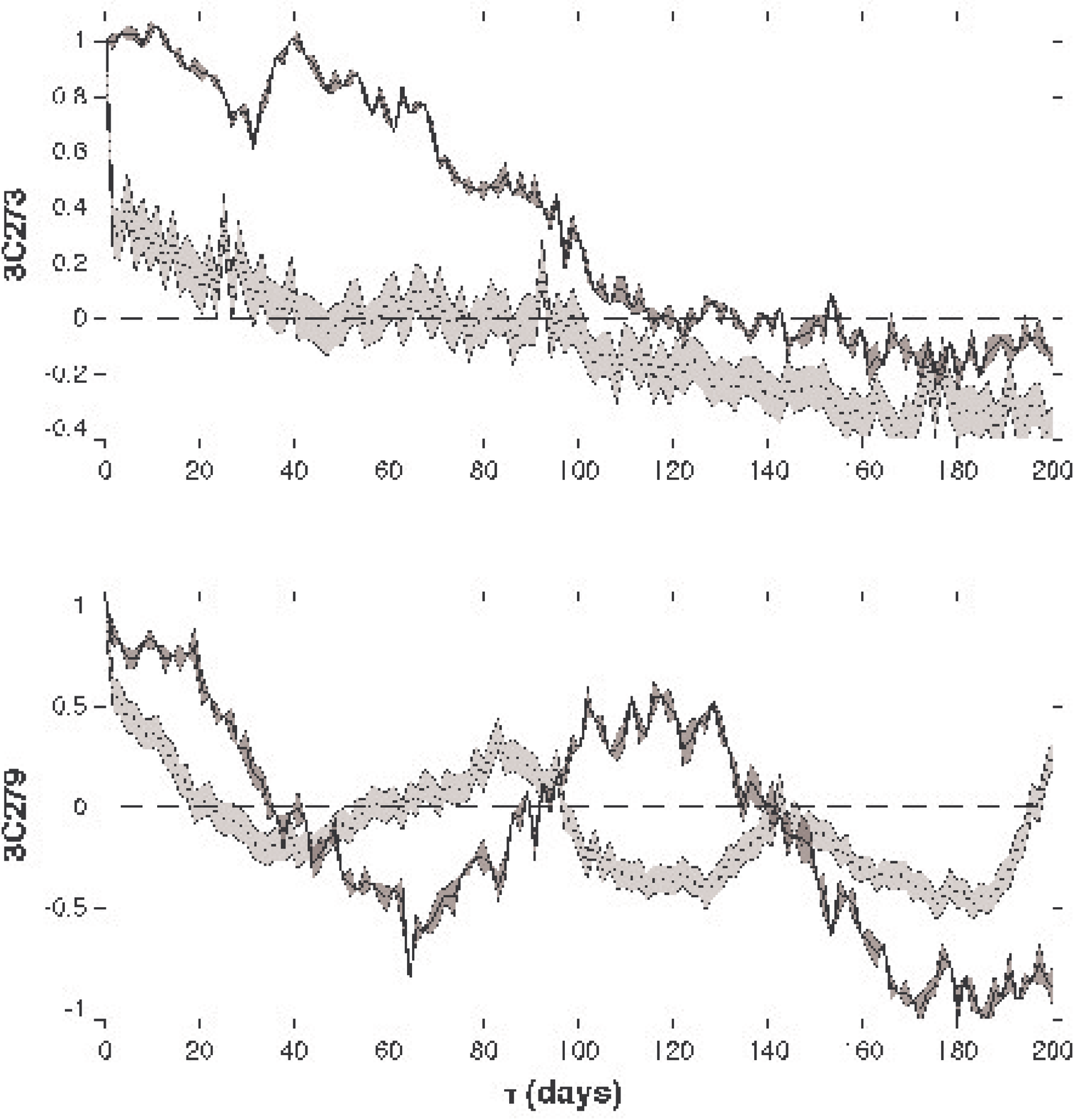}
\caption{Autocorrelation functions for the same
two AGN as in Figure 2 for radio and $\gamma$-ray data.
Solid line with dark error band: OVRO 15 GhZ; 
Dotted line with light error band: Fermi LAT.}
\end{figure}

Auto- and cross- correlation involving 
photon event data is a simple matter of
inserting the quantity in eq. (\ref{reciprocal})
into eq. (\ref{e_and_k}).
Since essentially any time series data mode
yields at least surrogates for $t_{n}$
and $x_{n}$, the same is true in general.
Figure 4 shows autocorrelation
functions computed in this way,
for the same AGNs shown in Figure 2 % \ref{bb_fig},
and Figure 5 
shows the corresponding cross-correlation
functions.
\begin{figure}
\centering
\vspace{30pt}\label{cross_examples}
\includegraphics[width=8cm]{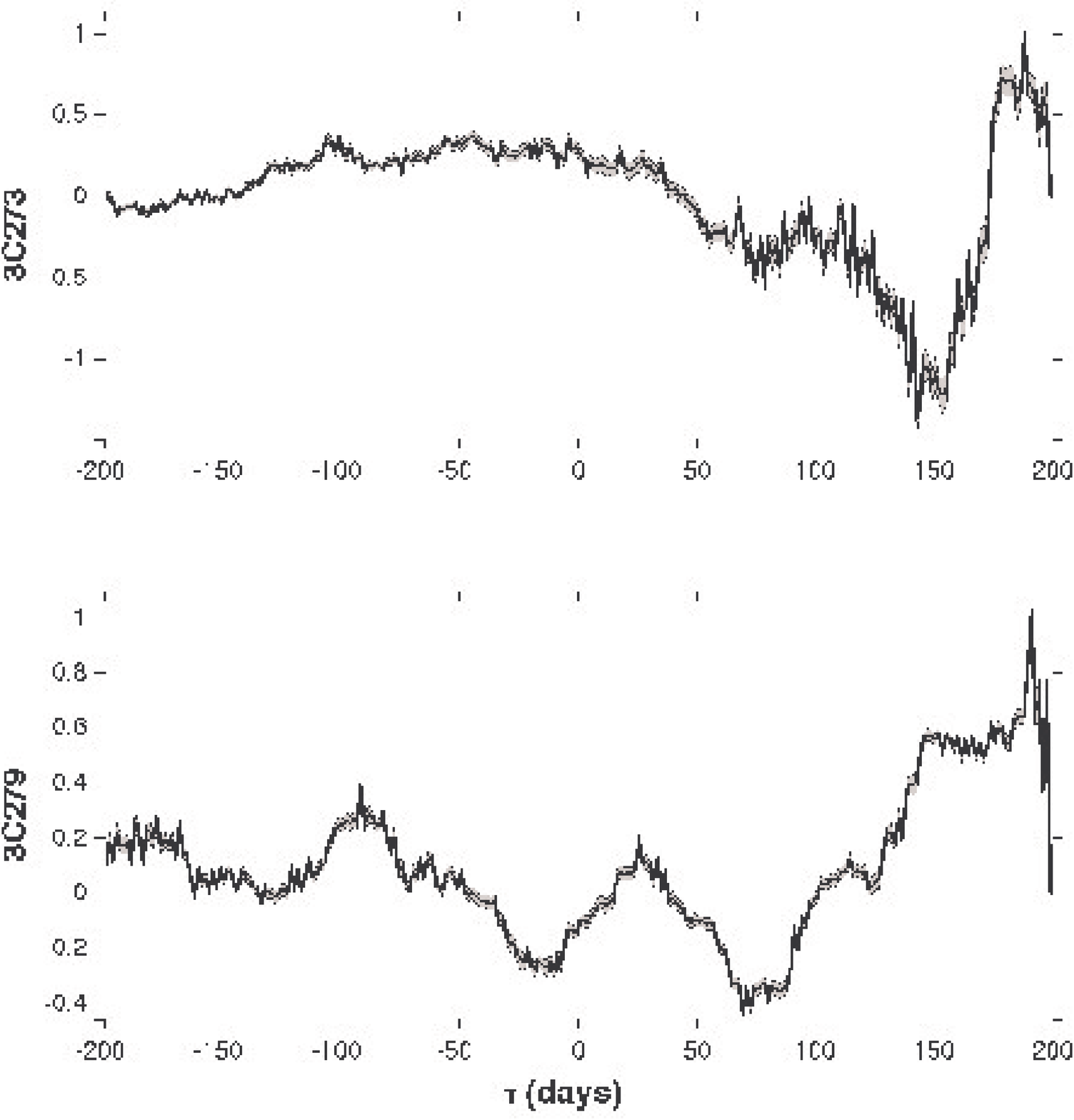}
\caption{Cross-correlation functions for the same
two AGN as in Figure 2, for radio and $\gamma$-ray data.}
\end{figure}

%____________________________________________________

\section{Fourier Power and Phase Spectra}
\label{power}

Perhaps the most used time series analysis
technique in astronomy is estimation of the
Fourier power spectrum, mainly with the
goal of detecting and then characterizing 
periodic signals hidden in noisy data,
but also for analyzing non-periodic signals 
such as quasi periodic oscillations and
colored, or ``${1 \over f}$,'' noise.
There are methods for direct 
estimation of Fourier power
(\cite{scargle_lomb}) and phase 
(\cite{scargle_ft})
spectra from time series data.
However, it is often more convenient 
to make use of the well-known
result that the power spectrum
is the Fourier transform of the ACF 
computed as described
above in \S \ref{auto_correlation}.
The sliding window power spectra
depicted in \S \ref{time_frequency} 
were computed in this way.
%__________________________________________________________

\section{Wavelet Representations}

It is relatively straightforward to
compute the wavelet transform
for any time series that can be
put into the standard data cell representation.
The wavelet shape (in this case the
piecewise constant Haar wavelet)
is integrated against the 
empirical signal amplitude 
assigned by the data cells.
Figure 6 shows the scalegrams,
or wavlet power spectra (\cite{drip}),
for the same AGN data as
in Figure 4.  There is not enough
data to yield much detail in these
spectral representations, but the
rough power law characteristic of
${1 \over f}$ processes can be seen, as well
as the noise floor for the LAT data.
\begin{figure}
\centering
\vspace{30pt}
%\special{psfile=fig_1.ps hscale=100 vscale=70100 hoffset=20 voffset=0}
\includegraphics[width=8cm]{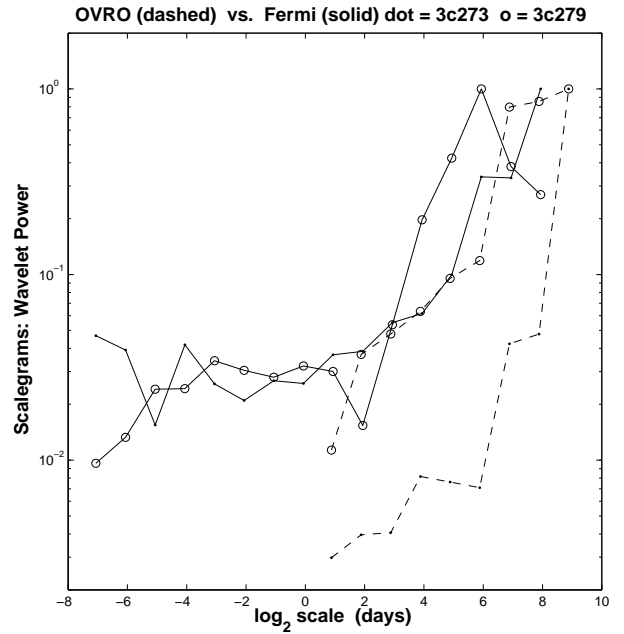}
\caption{Wavelet Power (scalegrams) for OVRO and LAT data 
on 3c273 and 3c279, with the Haar Wavelet.  
$log_{10}$ of the power plotted against $log_{2}$ of time scale in days.  }
\end{figure}

%__________________________________________________________
\section{Structure Functions}

Another concept in wide usage 
is the structure function. 
For the most part its auto- and cross-
versions are a repackaging
of the same information contained in 
the corresponding correlations.
This point has recently 
been emphasized by (\cite{uttley}).
In addition to summarizing
some of the caveats and problems
associated with structure functions,
these authors 
give a formal proof 
of the exact relation between
structure functions and
the corresponding auto- and 
cross-correlation functions.
In addition, the literature contains
a number of claims for the superiority
of the structure function that seem
unwarranted, especially in view of 
the relation just mentioned.
An example is the misconception 
that structure functions are somehow
immune from sampling effects, including
aliasing.  Finally, some analysts believe
that at short timescales the structure
function always becomes flat; the
actual generic behavior can be
derived from eq. (A10) of (\cite{uttley});
the normalized structure function satisfies 
\begin{equation}
NSF( \tau ) = 2 [ 1 - ACF( \tau ) ] \rightarrow C \tau ^{2}
\end{equation}
\noindent
for $\tau \rightarrow 0$, 
since autocorrelation functions are
even in $\tau$.
In practice this dependence may 
\emph{seem} flat compared to steeper 
behavior at intermediate time scales,
transitioning to the typical asymptotic
loss of correlation at large time scale
expressed as 
$NSF( \tau ) \rightarrow 2$,
correctly assessed
as flat.

A few other points perhaps favor 
the use of structure functions 
(beyond the fact that they have been
widely used in the past, and therefore
arguably should be computed if only for comparison
with previous work).
When the structure and correlation 
functions are estimated from actual
data, this equivalence result quoted above does not
hold exactly.  There can in fact be 
significant departures from the 
theoretical relations in Appendix A
of (\cite{uttley}),
due to end effects always present for 
finite data streams.
In addition, when measuring
slope of powerlaw relationships
it can be slightly more convenient
to fit polynomials to the typical shape of a
structure function than to the corresponding
correlation function or power spectrum.

%__________________________________________________________
\section{Time-Frequency Distributions}
\label{time_frequency}

The term \emph{time-frequency distribution}
refers to techniques for studying the time-evolution
of the power spectrum of time series.
This concept must deal with the fact that
the spectrum is a property of the entire
time interval, so that estimating it locally in 
time results in the need for trading off 
time resolution against frequency resolution.
See (\cite{flandrin}) for a complete
exposition of these issues.

There are many algorithms 
for computing time-frequency distributions, 
but little has been done for the case of event data,
one exception being the approach described in (\cite{gal}).
Although there are advanced techniques 
based on the Wigner-Ville distribution,
Cohen's class of distribution, and others,
in many applications 
the sliding window power spectrum 
is of considerable use.  The idea is simple:
compute the power spectrum of a subsample 
of the data within a restricted time-interval,
small compared to the total interval.
Information on the time dependence results
from the fact that the window is slid along
the observation interval.  Information on 
frequency dependence is contained in 
the power spectrum.
The tradeoff of time- and frequency 
resolution is mediated by the length of the
time window: a short window yields
high time resolution and low spectral
resolution, and \emph{vice versa} for a long window.
Implementation of this approach
is straightforward through use of the techniques
in \S \ref{auto_correlation} and \ref{power}.

\begin{figure*}
\centering
\vspace{30pt}
\includegraphics[width=16cm]{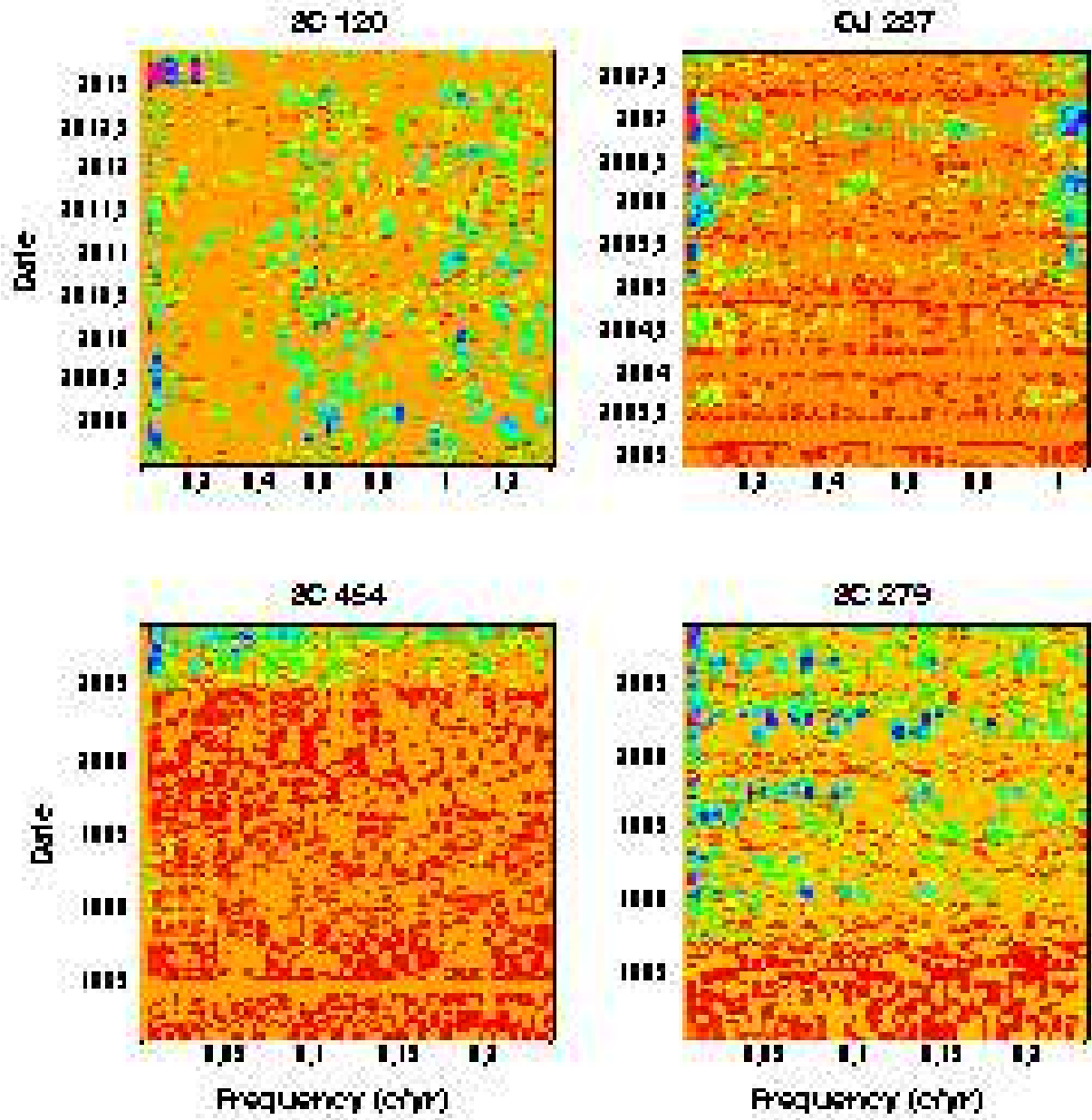}
\caption{Time-frequency distribution for
4 AGN data sets:
3c 120 x-ray data from Chatterjee et al. (2009, ApJ, 704, 1689)
provided by Alan Marscher;
optical, R magnitude data on OJ 278,
by Villforth C., Nilsson K., Heidt J., et al., 2010, MNRAS, 402, 2087,
provided by Ivan Agudo,
37 GhZ observations of 3c 454 and 3c 279  from the
Metsahovi Radio Observatory, provided by Anne Lahteenmaki.}
\end{figure*}
Figure 7 shows sliding window power spectra
computed, in this way, from time series data 
on four AGN provided by other authors 
at this workshop.  
These time-frequency distributions 
can show spectral details that are washed out 
in a power spectrum of the whole interval.
In these cases there is little evidence for
periodicities of any kind.
Note that these are preliminary results,
with no attempt being made to 
adjust the size of the window.

%__________________________________________________________
\section{Conclusions}

Rather than regurgitating the discussion above,
I end with a few practical suggestions.
They may seem obvious or trivial, but I have
found them surprisingly useful in practice.

When addressing time series data
in the form of eq. (\ref{data_intro}), 
the first step should be to study the
time intervals $t_{n+1} - t_{n}$;
in particular compute, plot, 
and study the their distribution with
suitably constructed histograms.
(Even if the provider of the data swears 
the times are evenly spaced, check it!)
This often reveals many defects in the data, such as
duplicate entries and observations out of order.
The outliers of the distribution signal peculiarities,
perhaps expected (such as known 
sampling irregularities, regularities, or semi-regularities)
but often unexpected surprises.
Figure 8
shows examples from the data for
which time-frequency 
distributions were shown above.
\begin{figure}
\label{dt_hist}
\centering
\vspace{30pt}
\includegraphics[width=8cm]{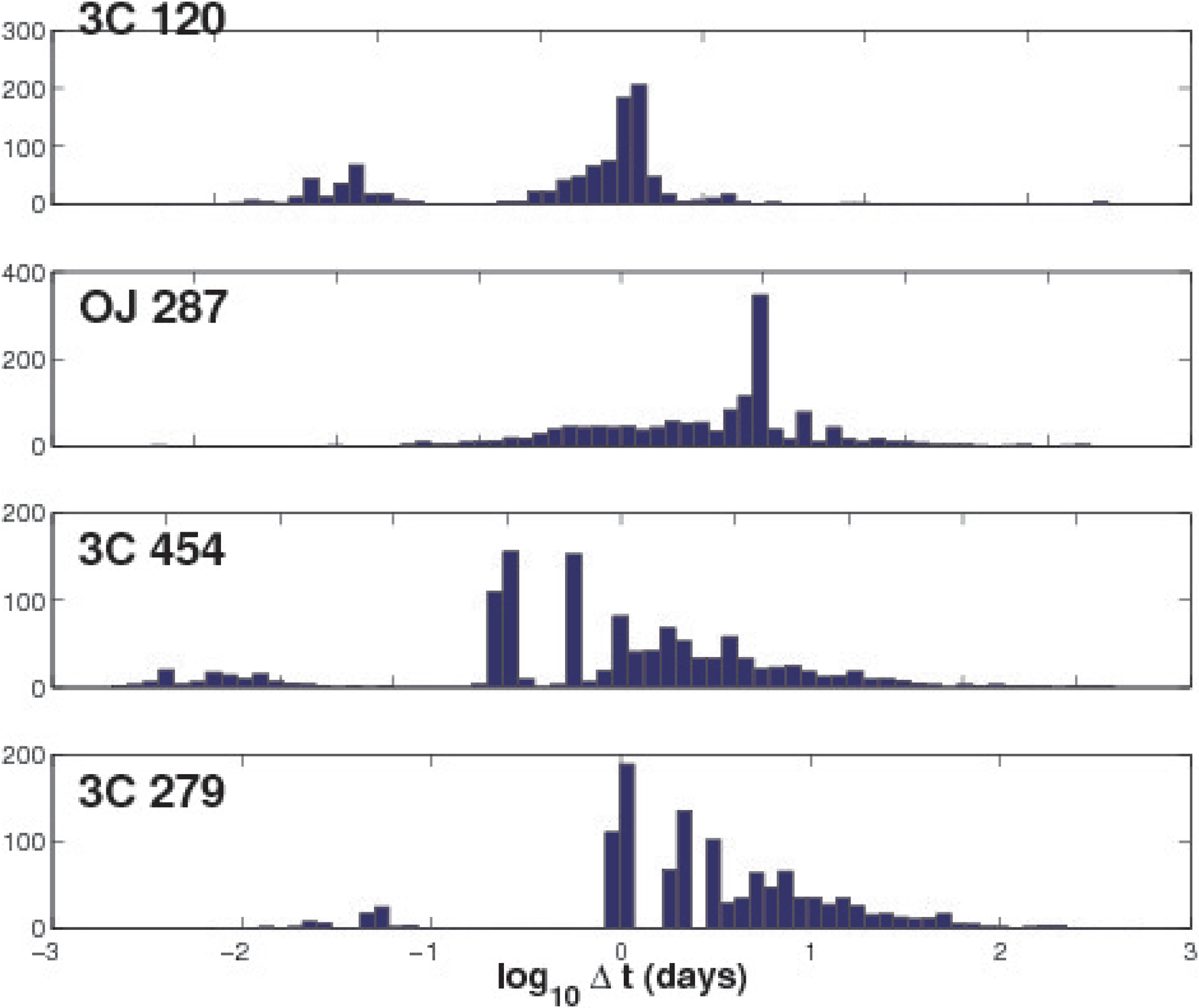}
\caption{Sampling histograms: the distributions
of the time intervals between the samples
for the data in Figure 7.}
\end{figure}
The reader is invited to see what 
conclusions can be deduced from these distributions.

Don't subtract the mean value!  Or at least
do so with attention to its effects.  Too often 
time series data are detrended without
careful consideration of the resulting effects on
the estimated functions.  Mean removal
is a special case of detrending.

While there are some cases where the
distinction between stationary and non-stationary
processes is important, with limited data it is
difficult or impossible to make this distinction
in practice.  For different reasons, the distinction between
linearity and non-linearity is best left to the
realm of physical models rather than data analysis.
Linearity is a property of physical processes,
and mathematical definitions (\cite{priestly,tong})
may or may not connect meaningfully to physical
concepts.

Finally, in thinking about AGN 
variability in general it is useful to
think in terms of the mathematical
concept of doubly stochastic
(or Cox) processes.  Essentially, this
is a picture in which there are two
distinct random processes: the intrinsic
variability of the source (truly random,
periodic, quasi-periodic, \emph{etc}.)
and the observation process. The latter
is random due to observational errors
from photon counting, detector noise,
background variability, \emph{etc.}
It is a major data analysis challenge
to cleanly separate out the observational
process to reveal the true variability
of the astronomical source.

\begin{acknowledgements}
For various contributions 
I am indebted to Brad Jackson, 
many members of the Fermi
Gamma Ray Space Telescope Collaboration,
especially 
Jay Norris,
Jim Chiang, and
Roger Blandford, 
and Tony Readhead,
Joey Richards, Walter Max-Moerbeck
and others in 
the Caltech Owens Valley Radio Observatory group, 
and to Alan Marscher,
Ivan Agudo,
Anne Lahteenmaki,
and
Sascha Trippe 
for kindly providing data sets.
\end{acknowledgements}

%\newpage

\end{document}